\documentclass[aps, twocolumn]{revtex4}
\usepackage{bm}
\usepackage{amssymb}
\usepackage{amsmath}
\usepackage{graphicx}

\begin{document}

\title{General Phase-Field Model with Stability Requirements on Interfaces in $N$-Dimensional Phase-Field Space}
\author{E. Pogorelov}
\email{evgeny.pogorelov@uni-bayreuth.de}
\author{J. Kundin}
\author{H. Emmerich}
\affiliation{Material and Process Simulation (MPS), University Bayreuth, 95448 Bayreuth, Germany}

\begin{abstract}
In this paper a general multi-phase-field model is presented which is an extension and modification of the model proposed by Folch and Plapp for three phase fields [R. Folch and M. Plapp,  Phys. Rev. E 72 011602 (2005)] to an arbitrary number of phases. In the model a physical constraint requiring  that the sum of all phase fields in the system is equal to one is resolved by the method of Lagrange multipliers. In fact, the thermodynamic driving force is reduced to its projection on the plane of the constraint. The general model functions in a $N$-dimensional phase-field space are derived which justify the requirements for the stability of the total free energy functional on dual interfaces and hence the absence of "ghost" phases. Furthermore, the case of the different interface energies and mobility parameters on the individual interfaces is resolved in a comprehensive manner. It is shown that the static equilibrium fulfils Young's law for contact angles with good accuracy. Then the 
model is verified by the quantitative simulation of the solidification in an Al-Cu-Ni alloy in the case of the four-phase transformation reaction. As a result, we found the way to control the dynamic of new phase nucleation using thermal noise in free energy functional.
\end{abstract}

\maketitle

\section{Introduction}
Multi-phase-field approaches developed in the recent years were practically applied to the simulation of the three-phase transformation in eutectic and 
peritectic alloys. The wide range of such realistic microstructure  studies was reported (see the review article \cite{Singer08} and the references therein). 
However, the investigation of four- and multi-phase transformation reactions is not fully covered.
One of the reason is the complexity of the models and the emerging challenges in terms of the understanding of multi-phase interactions.

In the last decade, two  main concepts of the multi-phase modelling have been established.
The first one is the multi-phase concept of Steinbach, 
which considers the change of any phase $i$ as the sum of other phase contributions due to the interaction between the phase $i$ and other phases on all individual interfaces, whereas the interaction between more than two phases are neglected. Therefore, the kinetic of individual interfaces can be considered separately with different interface energies and mobility parameters \cite{Steinbach96}. The physical constraint on phase fields, that the sum of all phase fields is equal to one, is kept in this model automatically.
The main works in this area are \cite{Tiaden98, Eiken06, Kim04}, where the main focus is the coupling of the kinetic equations to the diffusion equations in the multi-component systems.

The second concept supposes that the physical constraint on phase fields can be resolved by the formal application of the method of Lagrange multipliers to the free-energy functional, it works as the geometric projection of the driving force vector onto the plane of constraint.
The main representatives of this concept are Folch and Plapp\cite{Folch05}. They developed a qualitative phase-field model for three phases with a smooth designed model function in the free energy functional that should ensure the stability of the solution and the absence of "ghost" phases on the interfaces between two phases.
In this model the equations of motion for each interface can be mapped to the standard phase-field model of pure substances, where the thin-interface asymptotic analysis was applied \cite{Karma98}.
Moreover, it exploits the idea of second order expansion of the free energy functional\cite{Greenwood08,Echebarria04}, that simplifies the structure of the phase-field equations.
Later, in the work of Kundin and Siquieri \cite{Kundin11}, the total free energy functional proposed by Folch and Plapp was extended by using different thermodynamic factors of phases, whereas the evolution of the phase field was modeled according to the multi-phase concept of Steinbach.
Then the model was refined by the inclusion of cross terms in the thermodynamic factor matrix in multi-component systems \cite{Kundin12b}. The same authors \cite{Kundin12} applied the original model of Folch and Plapp to investigate the kinetics and morphology of the eutectic growth, where the difference in the thermodynamic factors was taken into account, too. They also shown that the nucleation of new lamellae on the boundary of the partner solid phase is rather an physical phenomena than an ``artifact'' of the model and the probability of this  nucleation is defined by the the undercooling and the surface tension that is in agreement with the classical nucleation theory.

The second concept was further exploited in the field of multicomponent alloy solidification \cite{Garcke04, Nestler05}. 
The authors used the Lagrange multiplier method and wrote the model in common form leaving out of consideration the stability requirements.
To prevent the existence of the third phase on the individual interfaces the authors added the third order term to the obstacle potential which hardly can be applied in more general cases.
Moreover, no thin-interface analysis of such models is available.

An intermediate approach between the first and second concept was proposed, which combines the different interface kinetics with the formal account for the change of the all phases in the multiple junctions \cite{Steinbach99}.
In the study \cite{Guo10} it was shown that the reformulated model predicts the angles between the phases close to the analytical values except of small deviation in the 3D simulations.
But by that approach the authors cannot overcome the criticism of Folch and Plapp about the instability of the solution for the chosen model functions \cite{Folch05}.

In this paper  we use the method of Lagrange multipliers and the idea of flatness and stability for model functions suggested by Folch and Plapp \cite{Folch05} and then extend it to the $N$-dimensional phase-field model (see Section \ref{Section_Formulation}). We construct the phase-field model functions step by step
and show the implementation of different interface energies and  mobility parameters for each individual interface. The numerical tests presented in Section \ref{Test_Young} were carried out to show how the model fulfills Young's law and how the mobility parameters influence the dynamic evolution of the corresponding individual interfaces. In Section \ref{Al-Cu-Ni} we verify the model by the qualitative simulations of the solidification in an Al-Cu-Ni alloy with a four-phase transformation reaction. In this section we explained the physics of four-phase solidification and write a full system of the model equations. We present the numerical simulations of the microstructure evolution and show the nucleation effects which can be observed. Finally, in conclusion we shortly summarize the main result of this paper.

\section{Formulation of general phase-field model}\label{Section_Formulation}
\subsection{Evolution equation for phase fields}

We assume that our physical system can be fully described by $N$ phase fields $ p_{i}\in[0,1]$, $i=1\dots N$ and by $n$ concentration fields $c^A\in[0,1]$, $A=1\dots n$. We identify a vector of phase fields as $\bm{p}=(p_{1},\dots,p_{N})$, where every phase field $p_i$ means the volume fraction of $i$-th phase. Therefore, we demand that in every time our system should follow the physical constraint, that is the sum of all phase fields should be equal to one
\begin{equation}
\sum_{i=1}^{N}p_{i}=1.
\end{equation}

The total free energy functional of the system  is written as
\begin{equation}
F=\int_{V}f\, dV,
\end{equation}
where a total free energy density is expanded into the following terms
\begin{equation}
f(\bm{p},\bm{\nabla p},\bm{c},T)=Kf_{g}(\bm{\nabla p})+Hf_{b}(\bm{p})+f_{c}(\bm{p},\bm{c},T).
\end{equation}
Here, $f_{g}$ sets a free energy cost depending on gradients of phase fields, forcing interfaces to have finite width. A constant $K$ has the dimension of energy per unit length and a constant $H$ has the dimension of energy per volume.
$f_{b}$ is a dimensionless barrier function, which is analog to the double well potential in the theory for two phases.
$f_c$ has the dimension of energy per volume and is the chemical part of the free energy which depends on concentration vector $\bm{c}=(c^{A},c^B\dots)$ and the temperature $T$.

For the evolution equations we chose Model C according to the classification given in Ref.~\cite{HalperinPRB1974, HalperinRevModPhys1977}. We used the Ginzburg-Landau equation for non-conserved field modified by Lagrange multiplier and the diffusion equation for conserved field
\begin{equation}
\begin{aligned}
&\tau(\bm{p})\frac{\partial p_{i}}{\partial t}=-\frac{1}{H}\frac{\delta F}{\delta p_{i}}\biggm|_{\sum_{j}p_{j}=1}\\
&=-\frac{1}{H}\biggl(\frac{\delta F}{\delta p_{i}}-\frac{1}{N}\sum_{j}\frac{\delta F}{\delta p_{j}}\biggr),\;i=1,\dots N,\\
&\frac{\partial \bm{c}}{\partial t} = \nabla\biggl[\mathbf{\hat{M}}(\bm{p})\nabla\frac{\delta F}{\delta \bm{c}}-\bm{J}_{at}(\bm{p})\biggr].
\end{aligned}
\label{EqLagrangeMulti}
\end{equation}
Here, $\tau(\bm{p})$ is a system relaxation time depending on the phase fields, $\mathbf{\hat{M}}(\bm{p})$ is a mobility matrix and $\bm{J}_{at}(\bm{p})$ is the anti-trapping current.

By using Lagrange multiplier method one get the projection of driving force onto the Gibbs simplex $S=\{\sum_{i}p_{i}=1\, p_{i}\in[0,1]\}$.
This automatically ensures $\sum_i \partial p_i/\partial t = 0$.

\subsection{Construction of model functions with stability and flatness requirement}

The first formulation of stability and flatness requirements for free energy model functions was made by Folch and Plapp in \cite{Folch05} as
\begin{equation}
\frac{\delta^2 F}{\delta p_k^2}\biggm|_{\sum_i p_i=1,p_k=0,1}>0\;\forall k,
\end{equation}
and
\begin{equation}
\frac{\delta F}{\delta p_k}\biggm|_{\sum_i p_i=1,p_k=0,1}=0\;\forall k,
\end{equation}
respectively.
In the following, we will use requirements which are equivalent to Folch and Plapp \cite{Folch05,Wheeler92} in three dimensional phase-field space but weaker in general $N$-dimensional space. Namely, we require on all interfaces $I_{ij}=\{p_j=1-p_i,\,p_{k\ne i\ne j}=0\}$ the stability condition
\begin{equation}
\frac{\delta^2 F}{\delta p_k^2}\biggm|_{\bm{p}\in I_{ij}}>0\, \forall i,j,k,\label{Stability1}
\end{equation}
and the flatness condition
\begin{equation}
\frac{\delta F}{\delta p_k}\biggm|_{\bm{p}\in I_{ij}}=0\, \forall i,j,k, \label{Flatness1}
\end{equation}
and analogical conditions on vertexes $V_i=\{p_i=1,p_{k\ne i}=0\}$
\begin{gather}
\frac{\delta^2 F}{\delta p_k^2}\biggm|_{\bm{p}\in V_i}\geqslant 0\;\forall i,k;\label{Stability2}\\
\frac{\delta F}{\delta p_k}\biggm|_{\bm{p}\in V_i}=0\;\forall i,k.\label{Flatness2}
\end{gather}

We start the construction of model functions with choosing the free energy gradient term in the form
\begin{equation}
f_g(\nabla \bm{p})=\frac{1}{2}\sum_i (\nabla p_i)^2.\label{fg}
\end{equation}
It is easy to check that $f_g$ satisfies the flatness conditions (\ref{Flatness1}),(\ref{Flatness2}). And the flatness requirement is the main reason why we are limited to such simple form (\ref{fg}).

Then, we construct the barrier function $f_b$ in such a way that we have an arbitrary positive interface energy $\sigma_{ij}$ for any individual  interface $I_{ij}$. Moreover, $f_b$ should satisfy our stability and flatness conditions (\ref{Stability1}-\ref{Flatness2}). For this aim we define a set of barrier functions
\begin{multline}
f_{b,ij}=\frac{1}{2}(z_{ij}|_{\phi=p_i}+z_{ij}|_{\phi=p_j}),\; z_{ij}=\phi^3(1-\phi)^3\\
-3(1-\phi)\phi^3\sum_{k\ne i,j}p_k^2+2\phi^3\sum_{k\ne i,j}p_k^3\,.
\end{multline}
It is the polynomial of minimum power which follows stability and flatness conditions and $f_{b,ij}(\bm{p})=0$ $\forall \bm{p}\in I_{kl}\ne I_{ij}$. Also if $\bm{p}\in I_{ij}$ and $p_i=\varphi$, $p_j=1-\varphi$, then $f_{b,ij}=\varphi^3(1-\varphi)^3$.
Finally, we can represent
\begin{equation}
f_b=\sum_{i<j}q_{ij}f_{b, ij}, \label{fb}
\end{equation}
where $q_{ij}$ will provide us the surface energy $\sigma_{ij}$ for each individual interface $I_{ij}$.

To determine constants $q_{ij}$ we should consider the evolution equation of phase fields (\ref{EqLagrangeMulti}) on an interface $I_{ij}$.
Using (\ref{Stability1}-\ref{Flatness2}) it can be shown that only $i$-th and $j$-th component of driving force are non zero. Then, taking into account that all terms $f_{b,kl}=0$ on $I_{ij}$ except $f_{b,ij}$, we will see that only $f_{b,ij}$ has a contribution in Eq.~(\ref{EqLagrangeMulti}). Analogically we can analyze the $f_g$ term. Therefore, without the chemical free energy term $f_c$  we can write for the static solution of Eq.~(\ref{EqLagrangeMulti})
\begin{gather}
\frac{\partial \varphi}{\partial t} = \frac{K}{H} \frac{\partial^2 \varphi}{\partial x^2}-q_{ij}\frac{\partial \varphi^3(1-\varphi)^3}{\partial \varphi}\rightarrow 0, \text{ for } t\rightarrow \infty\\
\Rightarrow\frac{\partial^2 \varphi}{\partial x^2}=\frac{q_{ij}H}{K}\frac{\partial \varphi^3(1-\varphi)^3}{\partial\varphi}\,.\label{EqStatic}
\end{gather}
In equilibrium the total free energy will turn to
\begin{equation}
F = \frac{K}{2}\biggl(\frac{\partial\varphi}{\partial x}\biggr)^2 + Hq_{ij} \varphi^3(1-\varphi)^3,
\end{equation}
where we use only one dimension for the sake of simplicity.
The solution of Eq.~(\ref{EqStatic}) can be expressed as
\begin{multline}
x(\varphi)=\int_{1/2}^\varphi \frac{\sqrt{K}}{\sqrt{2Hq_{ij}\phi^3(1-\phi)^3}}\, d\phi\,,\\
x\in(-\infty,\infty),\varphi\in[0,1].
\end{multline}
Then we have
\begin{multline}
\frac{\partial x}{\partial \varphi}=\frac{\sqrt{K}}{\sqrt{2Hq_{ij}\varphi^3(1-\varphi)^3}}\Rightarrow\\
\frac{\partial\varphi}{\partial x}=\sqrt{\frac{Hq_{ij}}{K}}\sqrt{2\varphi^3(1-\varphi)^3}
\end{multline}
and can express the surface energy as
\begin{multline}
\sigma_{ij}=\int_{-\infty}^\infty F\,dx=
\frac{K}{2}\int_0^1\frac{\partial\varphi}{\partial x}\,d\varphi\\
+Hq_{ij}\int_0^1\varphi^3(1-\varphi)^3
\frac{\partial x}{\partial\varphi}\,d\varphi\\
=\sqrt{HKq_{ij}} \int_0^1 \sqrt{2\varphi^3(1-\varphi)^3} \, d\varphi = a_1\sqrt{HKq_{ij}}.
\end{multline}
Then, $q_{ij}$ can be written as
\begin{equation}
q_{ij}=\frac{\sigma_{ij}^2}{HKa_1^2},\, \text{ where }
a_1=\frac{3\sqrt2}{128}\pi.
\end{equation}
and we rewrite Eq.~(\ref{fb}) as
\begin{equation}
f_b=\frac{1}{HKa_1^2}\sum_{i<j}\sigma_{ij}^2f_{b, ij}.
\end{equation}
Then constants $H$ and $K$ can be determined through a model interface width $W=\sqrt{K/H}$ and a maximal interface energy
$\sigma_{\max}=\max_{ij}\sigma_{ij}=a_1\sqrt{KH}$. That is $K=W\sigma_{\max}/a_1$ and $H=\sigma_{\max}/(Wa_1)$.
Using the above definitions we can write the barrier function as
\begin{equation}
f_b=\sum_{i<j}\frac{\sigma_{ij}^2}{\sigma_{\max}^2}f_{b,ij}.
\end{equation}
Therefore in our model we have different numerical interface widths $W_{ij}\sim 1/\sigma_{ij}$. If we change the interface energies $\sigma_{ij}$, then each particular interface width $W_{ij}$ will be changed automatically. We can also change the model interface width $W$ in accordance to the need of the numerical method.

Finally, we construct model functions $g_i(\bm{p})$ which will be used in the formulation of $f_c$. These functions should be equal to one on a vertex $V_i$ and 0 on other vertexes. They also should satisfy stability and flatness requirements (\ref{Stability1}-\ref{Flatness2}). Then we add a condition $g_i(0,\dots p_i,\dots,p_j,\dots 0)=1-g_j(0,\dots p_i,\dots,p_j,\dots 0)$ for $\bm{p}\in I_{ij}$, which comes from the thin-interface analyses.

We have found the model functions as polynomials of minimal power
\begin{multline}
g_i(\bm{p})=\frac{1}{2} p_i^2 \biggl( 15-25p_i+15p_i^2-3p_i^3\\
-15(1-p_i)\sum_{j\ne i} p_j^2\biggr).
\end{multline}
On all individual interfaces $I_{ij}$ the functions $g_i$ reduce to $p_i^3(10-15p_i + 6p_i^2)$ by taking into account the constraint $\sum_{k}p_k=1$.

To construct model functions $f_{b,ij}(\bm{p})$ (12) and $g_{i}(\bm{p})$ (23) in $N$ dimensional phase-field space we suggest to reduce the flatness (7-10) requirement to finite number of conditions. Therefore we assume that all our model functions should be respective symmetric polynomials of minimal power, where the symmetry is taken only such $p_k$, that $k\ne i$ for $g_i$ and $k\ne i,j$ for $f_{b,ij}$. We used the fundamental theorem for symmetric polynomial. That is, for every fixed power $S$ of polynomials $g_i$ and $f_{b,ij}$ we can represent them as finite expansion in terms of basic symmetric polynomials of power $s\leqslant S$. Due to the symmetry of respective derivatives we write Lagrange multiplier in a finite form. Then, from the flatness conditions (7-10) we get just a few equations instead of an undetermine number $N$.
Thus solving the linear system of equations for respective coefficients of expansion, we were able to find $g_i$ and $f_{b,ij}$ which satisfy all requirements (7-10).

\subsection{Evaluation of the smooth function for the mobility parameter}

Here, we propose a system mobility parameter $\tau^{-1}(\bm{p})$ which takes constant values $\tau_{ij}^{-1}$ on any individual interface $I_{ij}$ and smoothly varies on the Gibbs simplex $S$ and in neighborhood. The use of the mobility parameter $\tau^{-1}$ allows to consider any immobile interface $I_{ij}$ with a mobility parameter equal to zero $\tau^{-1}_{ij}=0$ instead of a relaxation time going to infinity $\tau_{ij} \rightarrow \infty$. 

Let us identify a distance $s_{ij}$ between a point inside the Gibbs simplex $S$ and an interface $I_{ij}$ as
\begin{eqnarray}
s_{ij}^2=\sum_{k\ne i,j} p_k^2+\frac{(p_i+p_j-1)^2}{2}.
\end{eqnarray}
Then we can write the mobility parameter of the system as a function of $s_{ij}$ in the form
\begin{eqnarray}
\tau^{-1}(\bm{p})=\sum_{i<j} \tau^{-1}_{ij} s^{-1}_{ij}/\sum_{i<j} s^{-1}_{ij}. \label{EqTau}
\end{eqnarray}
The result is plotted in Fig.~1, where it can be seen that the function (\ref{EqTau}) has prime lines on each dual interface as local minima.  It will preserves the stability in the vicinity of interfaces.

\begin{figure}
\begin{center}
\includegraphics[scale=0.5]{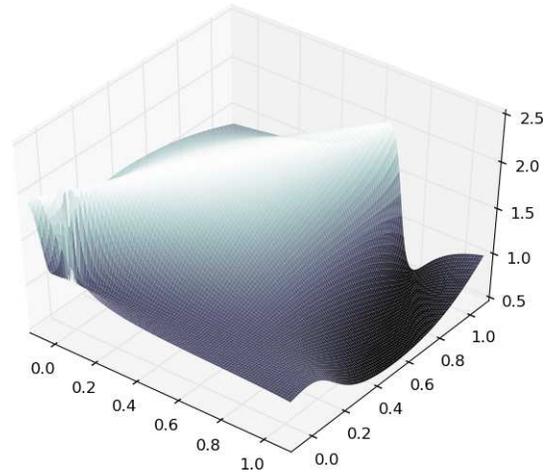}
\caption{The illustration plot of the mobility parameter$\tau^{-1}(p_1,p_2,p_3)$ on the plane $p_1+p_2+p_3=1$, where $\tau^{-1}_{12}=0.5$, $\tau^{-1}_{23}=1$, $\tau^{-1}_{13}=2$.}
\end{center}
\end{figure}

The alternative simpler formula, which provides the  similar simulation result, can be the following
\begin{equation}
\tau^{-1}(\bm{p})=\sum_{i<j} \tau^{-1}_{ij}p^2_i p^2_j/\sum_{i<j} p^2_i p^2_j\,.
\end{equation}

\subsection{Evolution equation for concentration fields coupling with equation for phase fields}
To evaluate the equations for the concentration fields we consider a multi-component system, which contains $n$ chemical components ($A$, $B$, \dots) excluding the solvent. We identify an equilibrium composition vector as $\mathbf{A}_{i}$ with components $A_{i}^A(T)$ and equilibrium chemical free energies of phases $B_{i}^A(\mathbf{A}_{i},T)=f_{c,i}(\mathbf{A}_{i},T)$. These parameters can be defined by the common plane construction to the free energy functions of individual phases. If there are more then one equilibrium composition for a phase $i$ with respect to all other phases $j\neq i$ we will take a mean value of them.

For the definition of the driving force of the phase transformation
we can write the mixture chemical free energy of a multi-component system as an interpolation between free energy functions of pure phases $f_{c,i}$ by using the second order Taylor expansion around the mixture equalibrium composition $c^{A,eq}=\sum_{i}A_{i}^Ag_{i}$
\begin{multline}
f_c=\sum_i B_i g_i+\sum_A^n \mu^{A,eq}\bigl(c^A-c^{A,eq}\bigr)\\
+\sum_{A,B}^n\frac{X^{AB}}{2}\bigl(c^A-c^{A,eq}\bigr)\bigl(c^B-c^{B,eq}\bigr).
\label{LagrangeMulti}
\end{multline}
Here $\mu^{A,eq}$ are the components of the equilibrium diffusion potential vector of the system. These parameters are important only for derivation of model, but are not included in final evolution equations. $X^{AB}$ are components of the mixture thermodynamic factor matrix which can be defined through the thermodynamic factor matrix of phases $\mathbf{\hat{X}_{i}}$ as
\begin{equation}
\mathbf{\hat{X}}^{-1}=\sum_{i}^N \mathbf{\hat{X}_{i}}^{-1}g_{i}.
\label{TF_Multi0}
\end{equation}
The derivation of this relation is given in \cite{Kundin12b}.

In the following we will use the mixture diffusion potential vector whose components are defined from the mixture chemical free energy as
\begin{equation}
\mu^A=\frac{\partial f_c}{\partial c^A}=\mu^{A,eq}+\sum_{B}^nX^{AB}\left(c^B-c^{B,eq}\right).  \label{TaylorMuMulti}
\end{equation}

The mixture chemical free energy~(\ref{LagrangeMulti}) gives the thermodynamic driving force of the phase transformation. The phase-field evolution equation can be written in an explicit form
\begin{multline}
\tau(\bm{p})\frac{\partial p_i}{\partial t}=
W^2 \biggl(\nabla^2 p_i-\frac{1}{N}\sum_k^N \nabla^2 p_k\biggr)\\
-\biggl(\frac{\partial f_b(\bm{p})}{\partial p_i}-\frac{1}{N}\sum_k^N \frac{\partial f_b(\bm{p})}{\partial p_k}\biggr)\\
+\frac{1}{H}\sum_j^N\frac{\partial g_{j}}{\partial p_i}\biggm|_{\sum_k p_k=1} \left(\sum_A^n\mu^{A}A_{j}^A-B_{j}\right).\label{PhasefieldEq0}
\end{multline}

Using the mixture diffusion potential vector the diffusion equations for all chemical components transform to the following form
\begin{equation}
\frac{\partial c^A}{\partial t}=\nabla \cdot \left[\sum_B^n M^{AB}(\bm{p})\nabla \mu^B-\bm{J}_{at}^A(\bm{p})\right],\label{B4}
\end{equation}
where $M^{AB}$ are the components of the mobility matrix $ \mathbf{\hat{M}}=\mathbf{\hat{D}}\cdot \mathbf{\hat{X}}^{-1}$. The components of the diffusion matrix are defined as $D^{AB}(\bm{p})=\sum_i^N D_i^{AB}g_i$, where $D_i^{AB}$ are the terms of the diffusion matrix in a phase $i$.
The values $\bm{J}_{at}^A$ are the anti-trapping currents for all components.
Then Eq.~(\ref{B4}) can be modified by the multiplication with $X^{AB}$ and the summation over all components as the equation in terms of the diffusion potential
\begin{multline}
\frac{\partial \mu^A}{\partial t}=\sum_B^n X^{AB}\nabla \cdot  \left[\sum_C^n M^{BC}\nabla \mu^C-\bm{J}_{at}^B(\bm{p})\right]\\
-\sum_B^n X^{AB}\sum_j^N\left(\frac{\partial g_{j}}{\partial t} A_{j}^B\right).\label{B5}
\end{multline}
Eqs. (30) and (32) are the evolution equations of the model.

\subsection{Evaluation of the derivatives for model functions $g_i$}

The full derivatives of the model functions $g_{i}$ according to (\ref{EqLagrangeMulti}a) are the following
\begin{gather}
\frac{\partial g_i}{\partial p_i}\biggm|_{\sum p_k=1}=\frac{\partial g_i}{\partial p_i}-\frac{1}{N}\sum_j\frac{\partial g_i}{\partial p_j},
\label{Deriv0}\\
\frac{\partial g_j}{\partial p_i}\biggm|_{\sum p_k=1}=\frac{\partial g_j}{\partial p_i}-\frac{1}{N}\sum_k\frac{\partial g_j}{\partial p_k},
\label{Deriv2}
\end{gather}
where
\begin{gather}
\frac{\partial g_i}{\partial p_i}=\frac{15p_i}{2}\Bigl((3p_i-2)\sum_{j\neq i} p_j^2-( 1-p_i)^2(p_i-2)\Bigr),\label{Deriv3a}\\
\frac{\partial g_i}{\partial p_j}=- 15p_i^2 (1-p_i)p_j,\label{Deriv3b}\\
\sum_{j\neq i}\frac{\partial g_i}{\partial p_j}=-15p_i^2(1-p_i)\sum_{j\neq i}p_j=-15p_i^2(1-p_i)^2. \label{Deriv3c}
\end{gather}
That yields
\begin{equation}
\frac{\partial g_{i}}{\partial p_i}\biggm|_{\sum p_k=1}= \frac{N-1}{N}\frac{\partial g_i}{\partial p_i}+ \frac{15}{N}p_i^2 (1-p_i)^2,
\end{equation}
\begin{multline*}
\frac{\partial g_{j}}{\partial p_i}\biggm|_{\sum p_k=1} = -\frac{1}{N}\frac{\partial g_j}{\partial p_j}+ \frac{15}{N}p_j^2 (1-p_j)^2\\
- 15p_j^2 (1-p_j)p_i,
\end{multline*}
After substitution of (\ref{Deriv3a}) and rearranging we have

\begin{multline}
\label{Deriv4}
\frac{\partial g_i}{\partial p_i}\biggm|_{\sum p_k=1}=\frac{15(N-1)}{2N}p_i
\Bigl((3p_i-2)\sum_{j\neq i} p_j^2\\
+(3p_i+2)(1-p_i^2)\Bigr)-\frac{15(2N-3)}{N} p_i^2 (1-p_i)^2,
\end{multline}
\begin{multline}
\label{Deriv5}
\frac{\partial g_j}{\partial p_i}\biggm|_{\sum p_k=1}=-\frac{15}{2N}p_j
\Bigl((3p_j-2)\sum_{k\neq j}p_k^2
+(3p_j+2)(1-p_j^2)\Bigr)\\
+15\frac{3}{N}p_j^2(1-p_j)^2-15p_j^2(1-p_j)p_i,
\end{multline}
Using Eqs.~(\ref{Deriv4}) and (\ref{Deriv5}) it can  be derived that for any individual interface $I_{ij}$ these derivatives are independent of $N$ and will be reduced to $15\, p_i^2 (1-p_i^2)$ and $-15\, p_j^2 (1-p_j^2)$, respectively.

\subsection{Relation to the previous model}

Let us notice that if two polynomials are equal to each other at the Gibbs simplex $S$, then they are equivalent in our model, because they have the same derivatives projected at $S$.
Keeping it in mind, we found that our functions $g_i$ are equivalent to the analogical functions suggested by Folch and Plapp in \cite{Folch05} for a 3-phase system. Moreover, the derivatives $\frac{\partial g_{i}}{\partial p_i}\vert_{\sum p_k=1}$ completely coincide with the derivatives of  functions $g_i$ in Ref. \cite{Folch05} .
The derivatives $\frac{\partial g_{j}}{\partial p_i}\vert_{\sum p_k=1}$ can be written in the form
\begin{multline}
\frac{\partial g_j}{\partial p_i}\biggm|_{\sum p_k=1}=-\frac{1}{(N-1)}
\frac{\partial g_j}{\partial p_j}\biggm|_{\sum p_k=1}\\
+\frac{15}{(N-1)}p_j^2(1-p_j)\Bigl(\sum_{k\neq i,j}p_k-(n-2)p_i\Bigr).
\end{multline}
For a three-phase system it will be reduced to
\begin{equation}
\frac{\partial g_j}{\partial p_i}\biggm|_{\sum p_k=1}=-\frac{1}{2}\frac{\partial g_j}{\partial p_j}\biggm|_{\sum p_k=1}\!\!\!\!\!\!\!\!
+\frac{15}{2}p_j^2(1-p_j)(p_k-p_i),
\end{equation}
and is similar to the derivatives of functions $g_j$ in Ref. \cite{Folch05}, too.
There is the typo in the Eq.~(3.25) of the Ref.~\cite{Folch05}, where instead of $\frac{\partial g_j}{\partial p_j}$ it is written $\frac{\partial g_i}{\partial p_i}$.
If one will use this equation with typo, then the wrong unexpected behavior of Folch and Plapp model can be observed. It can lead to the additional nucleation of a phase $p_{i\ne j,k}$ on the individual interface $I_{jk}$.

\section{Numerical tests to check Young's law, influence of mobility parameters, and absence of ghost phases}\label{Test_Young}

The phase-field model should asymptotically convert towards a sharp interface model when the thickness of interface goes to zero. The corresponding sharp-interface model should fulfill the Young's law. The aim of this section is to test how does our model fulfill the Young's law in the case of different interface energies and mobility parameters.

In all simulations we used the chemical free energy $f_c=0$ to test the influence of interface energies and mobility parameters and to test whether ``ghost'' phases exist or not. Here, all simulations were done in a two dimensional coordinate space having 128x128 discrete points with the fixed boundary conditions. During the tests we measured the position of the interfaces as a mixture of two phases and the position of triple points as the mixture of three phases. For simulations in 2D space we did not find points having the mixture of four phases. In Fig.~2 the initial states for all tests are shown.

The initial state of test 1 is shown in Fig.~2(a). We investigated the evolution of three phases determined in the three-dimensional phase-field space with similar interface energies and mobility parameters $\sigma_{12}=\sigma_{13}=\sigma_{23}$, $\tau_{12}=\tau_{13}=\tau_{23}$. Then we carried out test 2 for four different phases determined in the four-dimensional phase-field space with similar initial state and the same parameters Fig.~2(b).  The results of the simulation are shown in Fig.~3. We found that both systems have the same dynamic evolution with a good accuracy as it was expected. In the static equilibrium  we got in the three-phase junction angles of 120$^\circ$  with a high accuracy, which satisfies the Young's law. We can also see clearly in these figures that the numerical interface widths are constant and we do not have ``ghost'' phases on each individual interface.

In Fig.~4 the dynamic of triple points is shown for tests 1 and 2. We have found that interfaces (mixture of two phases) in these tests are very close to straight lines. In this figure we show numerical assymptotic limits which are very closed (within 0.1\% of simulation box) to the analytical limits  calculated by using Young's law.

\begin{figure}
\label{Fig2}
\begin{center}
\begin{tabular}{cc}
\includegraphics[scale=0.4]{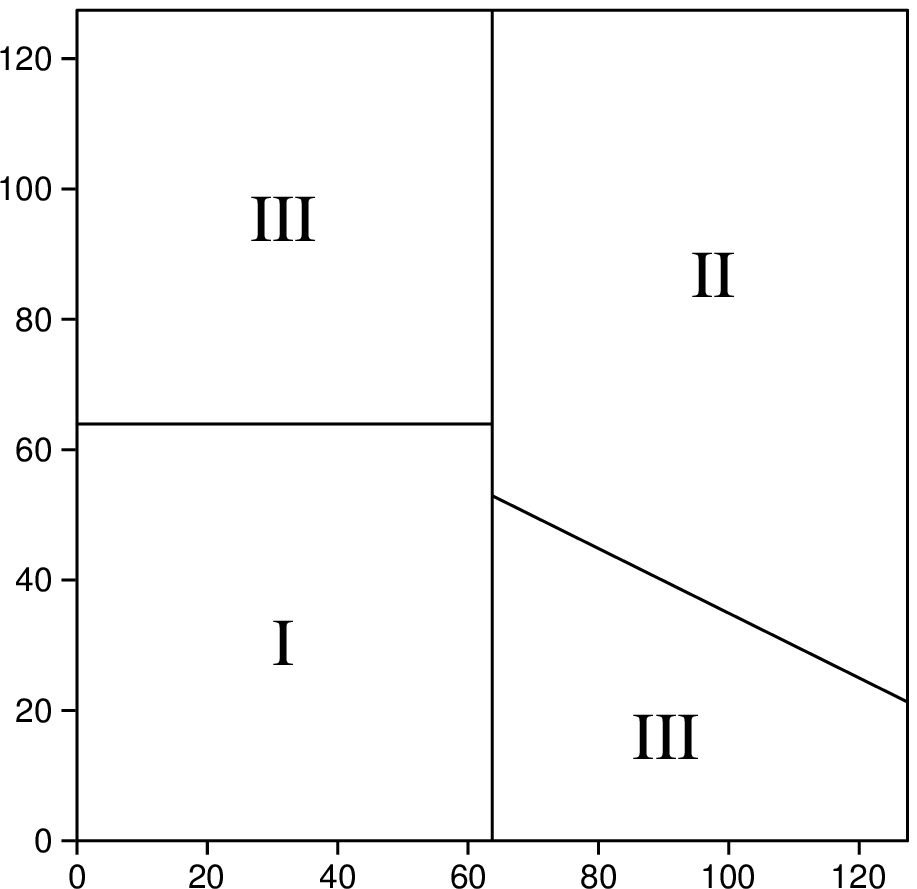} (a)&
\includegraphics[scale=0.4]{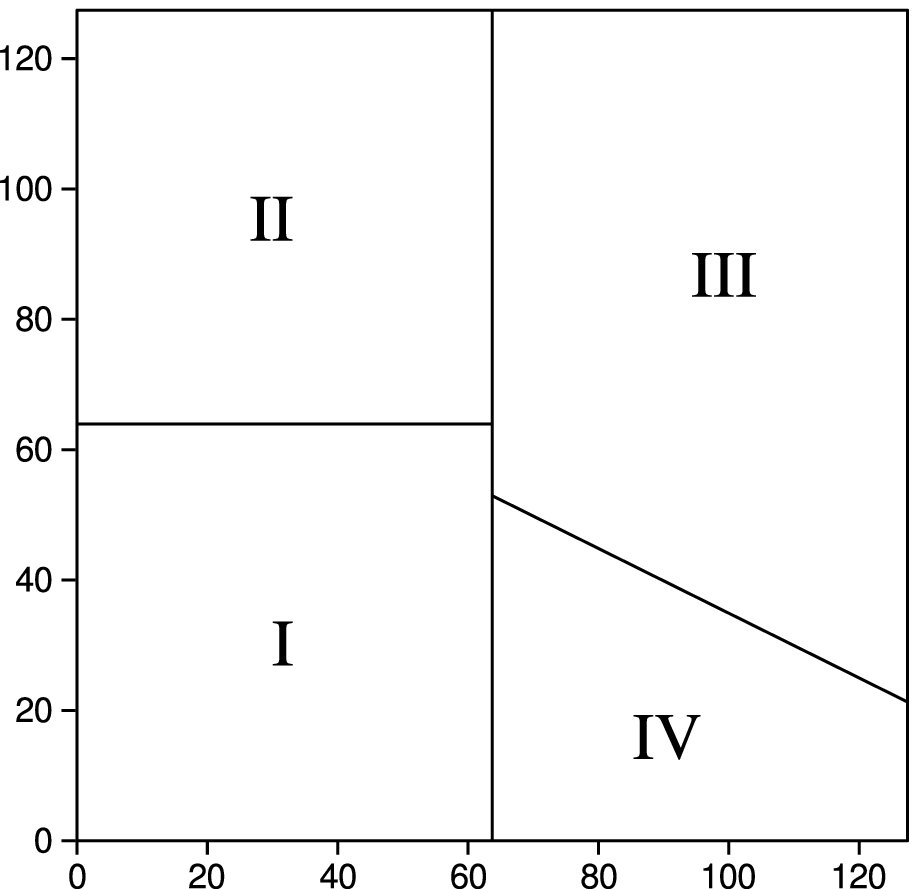} (b)\\
\includegraphics[scale=0.4]{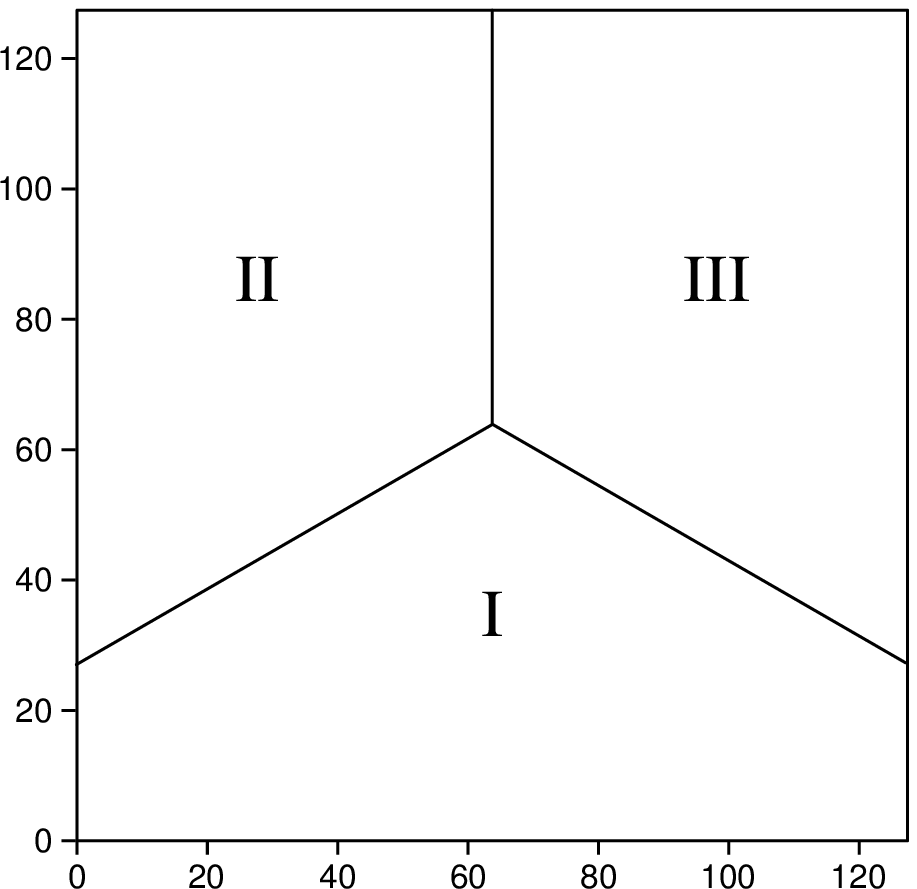} (c)&
\includegraphics[scale=0.4]{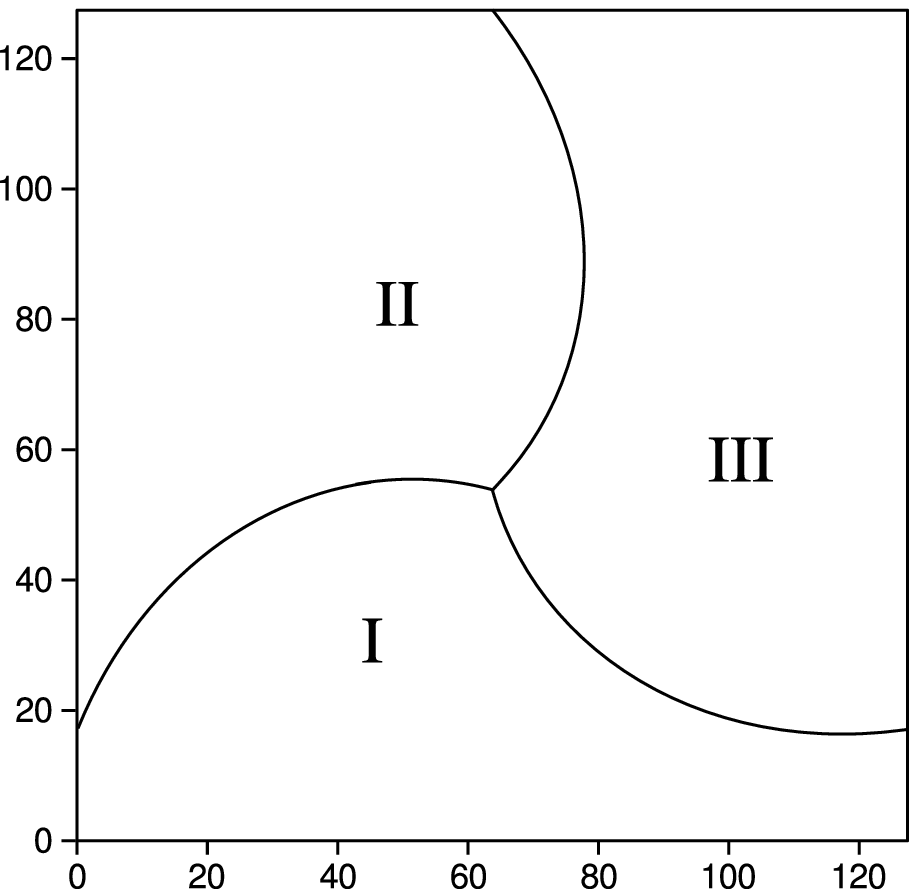} (d)
\end{tabular}
\caption{Initial configurations for numerical tests (128x128 discrete points). (a) --- test 1; (b) --- test 2, geometrically is equivalent to (a), but consists 4 phase fields; (c) --- test 3 and test 4, triple point is in the center of square, all angles are equal to $120^\circ$, (d) --- triple point has ($x_0$, $y_0$) coordinates, where $x_0=64$ and $y_0=128(1-1/\sqrt{3})$, interface curves in polar coordinates with center ($x_0$, $y_0$) have equations $r=r_0(4\varphi/\pi+\text{const})$ with initial angles $120^\circ$, where $r_0=128/\sqrt{3}$.}
\end{center}
\end{figure}

\begin{figure*}
\label{Fig3}
\begin{center}
\includegraphics[scale=0.4]{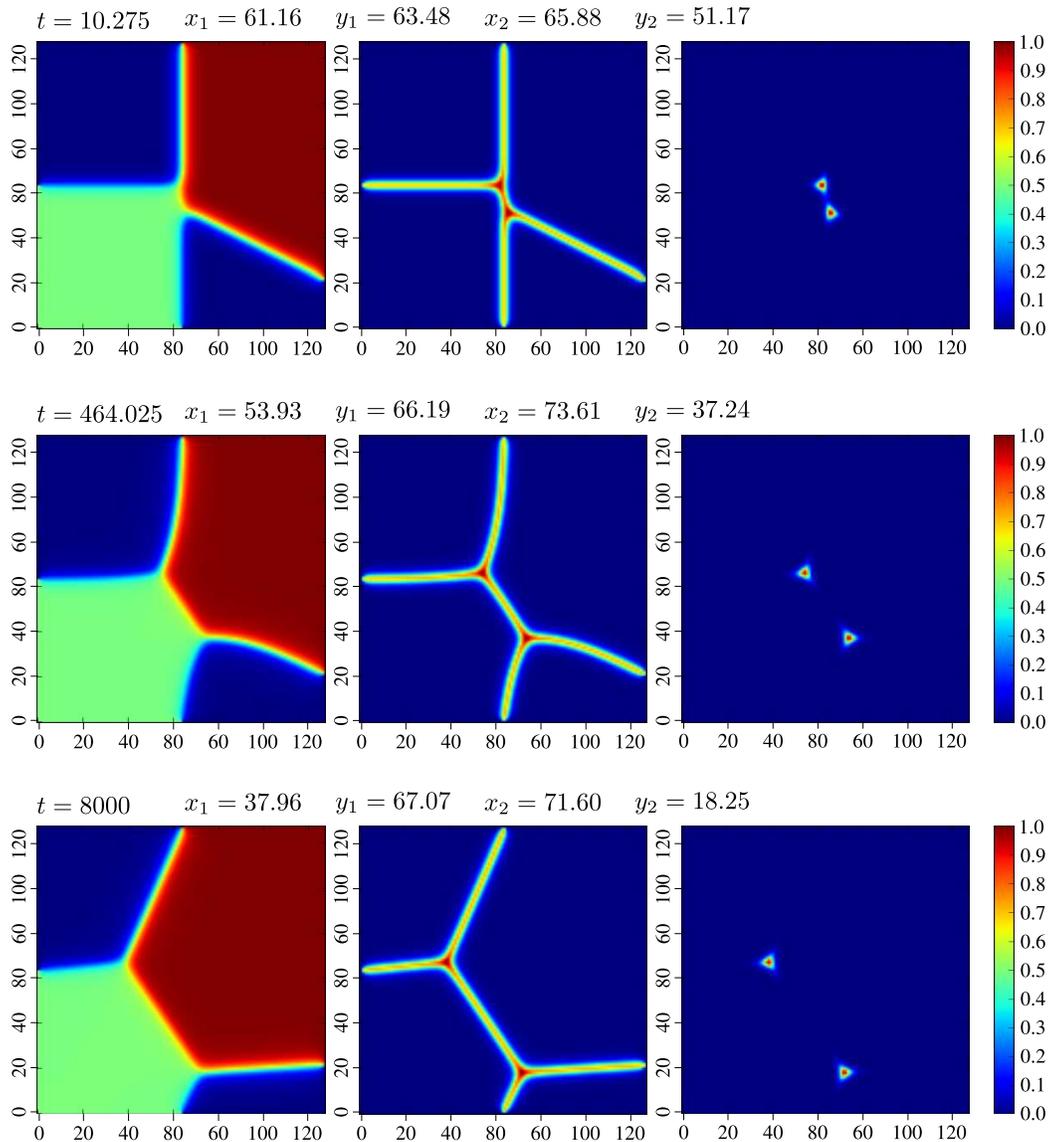} 
\caption{Test 1 and 2. The time evolution of the system having initial states presented in Figs.~2(a,b) with equal $\sigma_{12}=\sigma_{13}=\sigma_{23}$ and $\tau_{12}=\tau_{13}=\tau_{23}$. In the first column we show three and four different phases for test 1 and test 2, respectively (the difference during the entire simulation time is indistinguishable). In the second column we show the interfaces (mixture of two or three phases). In the third column we show triple points (mixture of three phases). There are no quadropole points in our simulation. In the first, second and third row we show the system in the initial ($t=10.3$), the intermediary ($t=464$) and the final state ($t=8000$).}
\end{center}
\end{figure*}

\begin{figure}
\begin{center}
\includegraphics[scale=0.35]{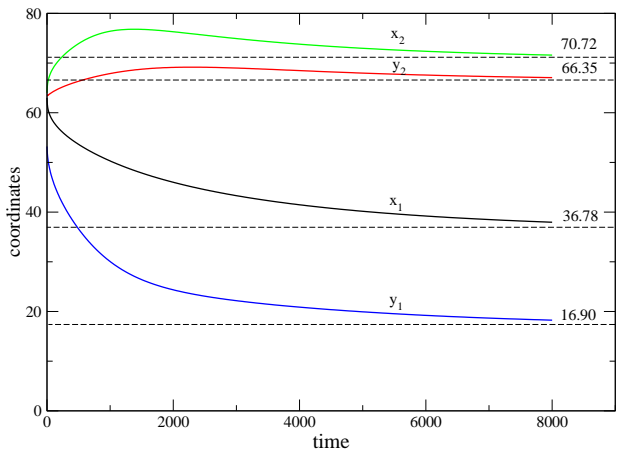} 
\caption{Test 1 and 2. The time evolution of the system having initial states presented in Figs.~2(a,b) with equal $\sigma_{12}=\sigma_{13}=\sigma_{23}$ and $\tau_{12}=\tau_{13}=\tau_{23}$. Here we show the evolution of triple point coordinates ($x_1$, $y_1$) and ($x_2$, $y_2$). Numerical assymptotic limits which are very closed (within 0.1\% of simulation box) to the analytical limits calculated by using Young's law.}
\end{center}
\end{figure}

In tests 3 and 4 we have simulated three different phases defined in a three dimensional phase-field space. The system has different interface energies with ratios $\sigma_{12}:\sigma_{13}:\sigma_{23}=0.5:0.75:1$. In test 3 we have equal mobility parameters $\tau_{12}=\tau_{13}=\tau_{23}$ and in test 4 we have different mobility parameters for each interface with the following ratios $\tau_{12}:\tau_{13}:\tau_{23}=2:1:(4/3)$. Therefore for tests 3 and 4 we have a different evolution of the systems, but the same static equilibrium state. In the initial state of both tests the angles between the interfaces in the triple point are $\gamma_{12}=\gamma_{13}=\gamma_{23}=120^\circ$ as shown in Fig.~2(c). According to Young's law the respective angles in the static equilibrium should follow the ratios $\sin\gamma_{12}:\sin\gamma_{13}:\sin\gamma_{23}=\sigma_{12}:\sigma_{13}:\sigma_{23}$. From this ratio we can calculate the equilibrium angles as $\gamma_{12}=151^\circ$, $\gamma_{13}=133.5^\circ$, $\gamma_{23}=75.5^\circ$.
Therefore we got different numerical interface widths $W_{ij}$, which are inverse proportional to the corresponding interface energies $\sigma_{ij}$, as it was expected. Test 4 is shown in Fig.~5 in a similar way to Fig.~3, except of a fourth column where the shifted mobility parameter $(\tau^{-1}(\bm{p})-0.5)/2$ is shown.  From this figure we can see that the Eq.~(\ref{EqTau}) gives us $\tau^{-1}(\bm{p})=\tau_{ij}^{-1}$ on every individual interface $I_{ij}$ with very good accuracy.

To check the difference in the evolution for tests 3 and 4 more precisily, we show the evolution of the triple point coordinates in Fig.~6 with numerical assymptotic limits which are very close (within 0.1\% of simulation box) to analytical limits calculated by using Young's law.

\begin{figure*}
\label{Fig4}
\begin{center}
\includegraphics[scale=0.3]{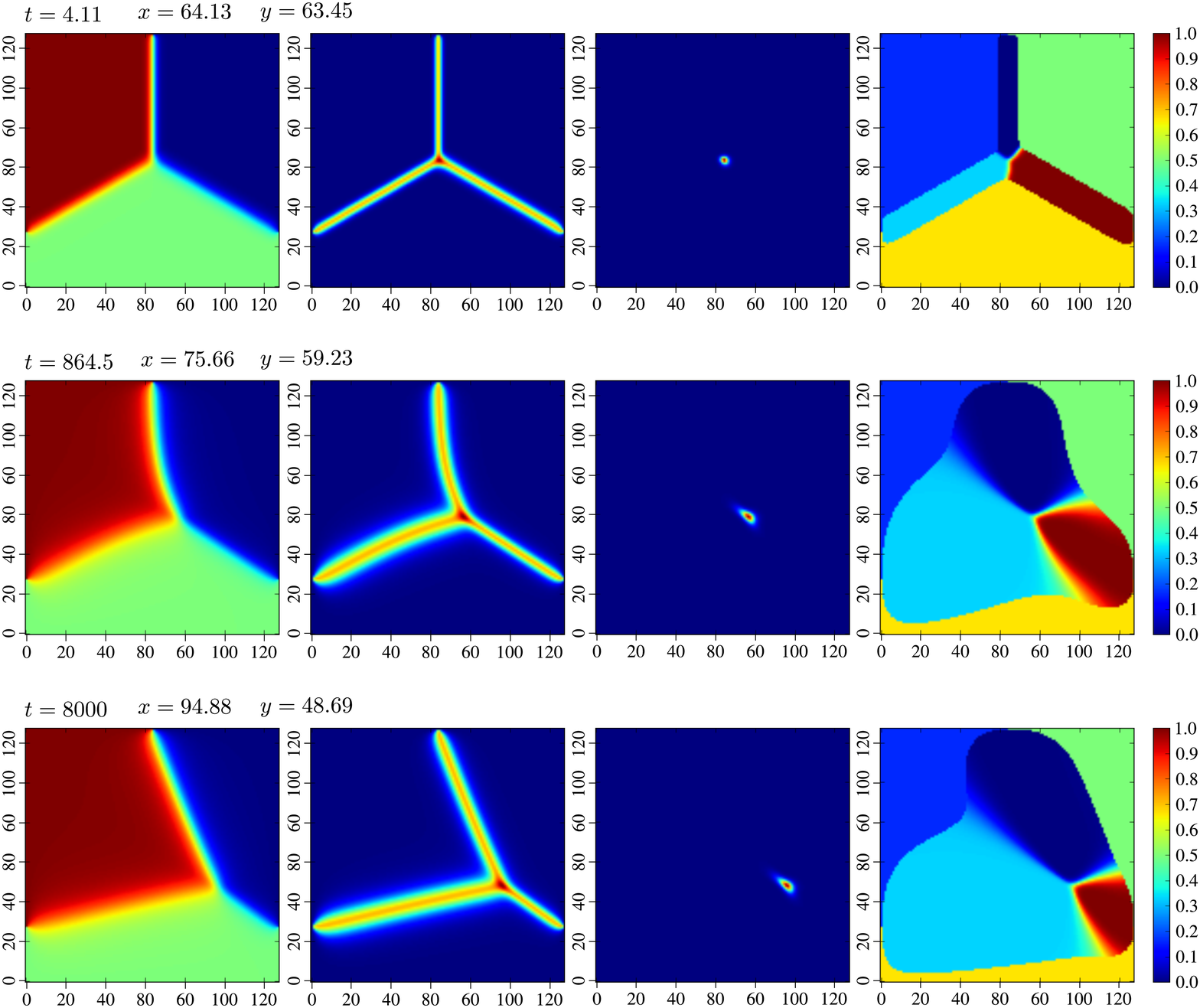} 
\caption{Test 4. The time evolution of the system presented in Fig.~2(c) with $\sigma_{12}:\sigma_{13}:\sigma_{23}=0.5:0.75:1$, $\tau_{12}:\tau_{13}:\tau_{23}=2:1:4/3$. In the first column three different phases are shown, in the second column --- a mixture of two or three phases, in the third column --- a mixture of three phases, in the fourth column --- the shifted mobility parameter is shown $(\tau^{-1}(\bm{p})-0.5)/2$. In the first, second and third row the system is at the initial, the intermediary and the final state.}
\end{center}
\end{figure*}

\begin{figure}
\label{Fig5}
\begin{center}
\includegraphics[scale=0.35]{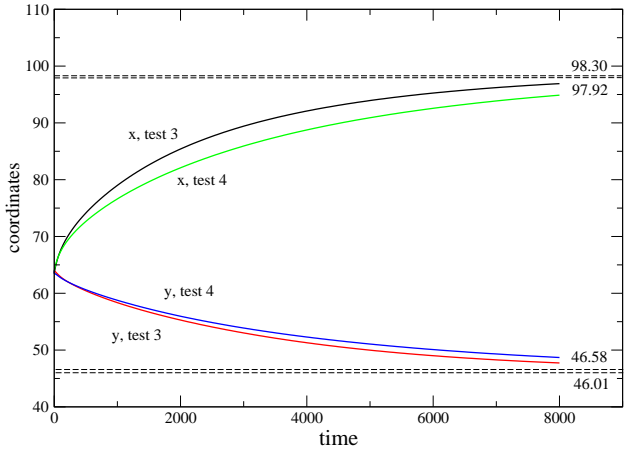} 
\caption{The time evolution of the triple point (mixture of three phases) for tests 3 and 4. We have shown numerical assymptotic limits (98.3, 46.01) and (97.92, 46.58) for tests 3 and 4, respectively, which are very close (within 0.1\% of simulation box) to the analytical limits calculated by using Young's law.}
\end{center}
\end{figure}

In tests 5, 6 (Figs.~7, 8) we checked the influence of mobility parameters on the evolution of the system. That is how quickly an interface will be straightened and what is the evolution of the triple point. The initial configuration is shown in Fig.~2(d). All initial interfaces have the central symmetry and the same length.
We used an equal interface energies $\sigma_{12}=\sigma_{13}=\sigma_{23}$ in both tests. For test 5 we used an equal inverse mobility $\tau_{12}=\tau_{13}=\tau_{23}$ and for test 6 we have $\tau_{12}:\tau_{13}:\tau_{23}=2:1:4/3$. Therefore in test 5 we got the evolution with the central symmetry in the triple point, whereas in test 6 the symmetry is broken as it was expected. We show the evolution of triple points for both tests in Fig.~7 and the deviance from straight lines for different interfaces in Fig.~8. We have shown that numerical assymptotic limits for both tests are very close (within 0.1\% of simulation box) to predicted values by the Young's law.

The triple point in test 5 have a small deviation from the initial configuration due to numerical errors. We also got similar deviance from straight lines for different interfaces $D_{12}=D_{13}=D_{23}$ with high accuracy in this test. The picture is different for test 6 due to assymetric mobility parameters.

\begin{figure}
\begin{center}
\includegraphics[scale=0.35]{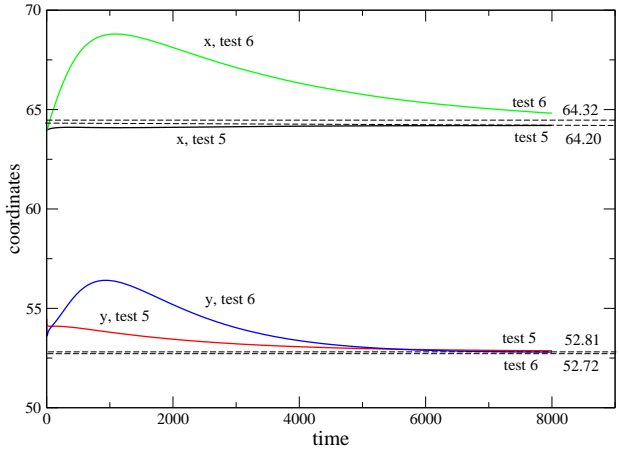} 
\caption{The evolution of the triple point coordinates for tests 5 and 6. We have shown numerical assymptotic limits (64.20, 52.81) and (64.32, 52.72) for tests 5 and 6, respectively, which are very close (within 0.1\% of simulation box) to the analytical limits calculated by using Young's law.}
\end{center}
\end{figure}

\begin{figure}
\label{Fig7}
\begin{center}
\includegraphics[scale=0.35]{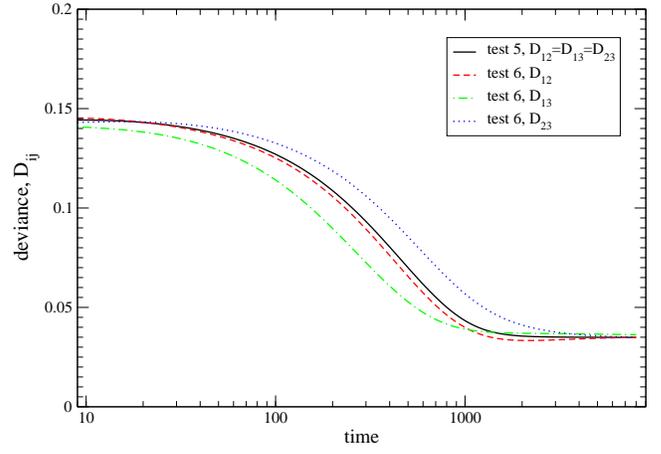}
\caption{The evolution of deviances $D_{ij}$ from straight lines of correspondent interfaces $I_{ij}$.}
\end{center}
\end{figure}

\section{Numerical test for a four-phase reaction}\label{Al-Cu-Ni}

\subsection{Alloy system and model parameters}

For the numerical test we have chosen a ternary Al-Cu-Ni alloy in the Al reach corner of the phase diagram.
A zoomed view of the Al reach corner in Fig.~9 shows the liquidus surface, the boundary curves and the regions where NiAl$_3$, Ni$_2$ Al$_3$ and (Al) phases solidify firstly, which we identify as $\alpha$, $\beta$ and $\gamma$ phases respectively. Between the $\alpha$ and $\beta$ 
phases there is the peritectic line $p_5$. Between $\alpha$ and $\gamma$, $\beta$ and $\gamma$ phases there are eutectic lines. The three-phase peritectic reaction ($p_5$) and four-phase reactions ($U_5$) and ($U_7$) are indicated.

\begin{figure}
\label{Fig8}
\begin{centering}
\includegraphics[scale=0.35]{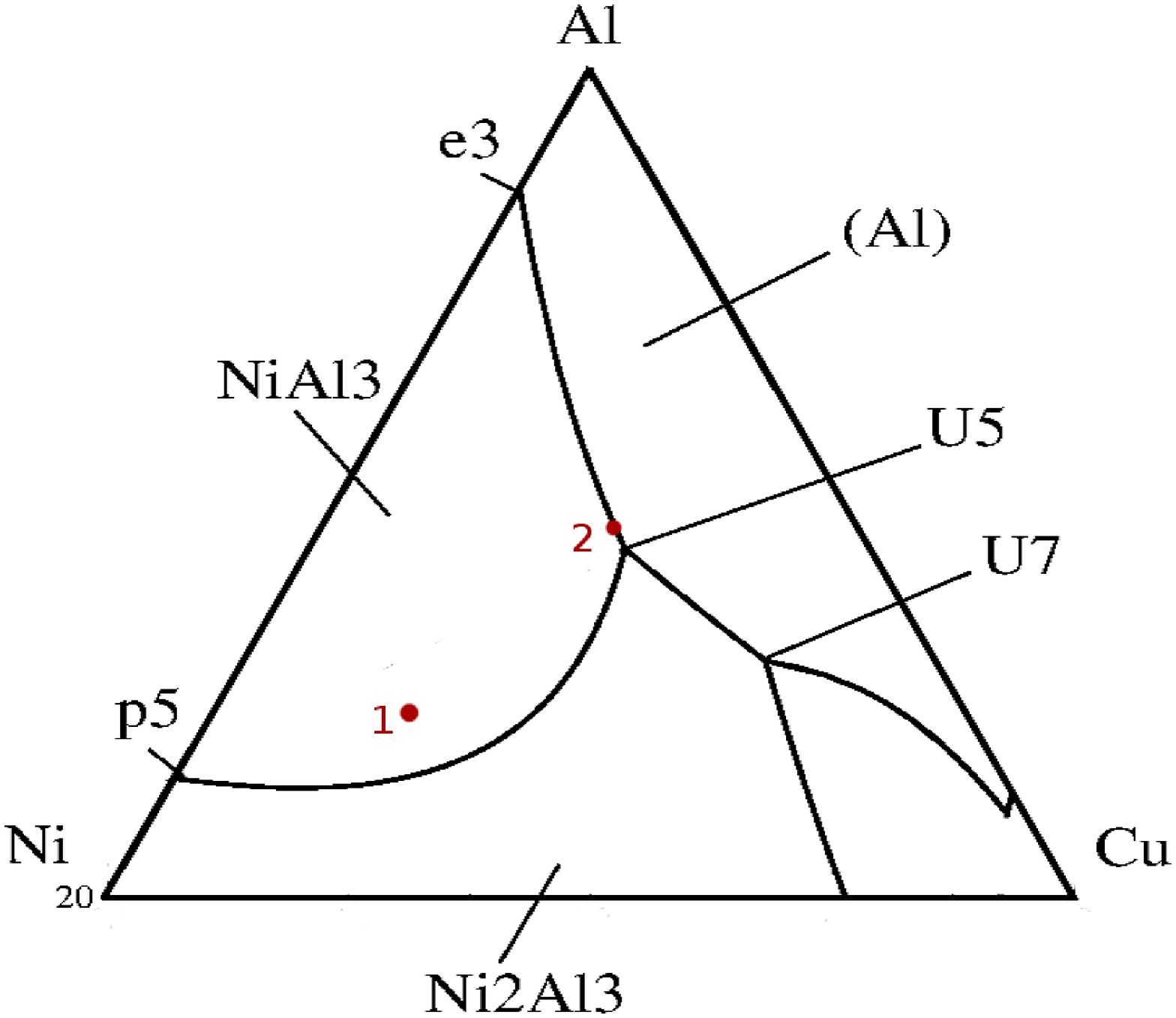}
\caption{
The liquidus surface  in Al-reach corner of Al-Cu-Ni phase diagram.}
\end{centering}
\end{figure}

As an example we consider the solidification of an alloys with the initial concentration of the liquid of 11 at$\%$Ni - 4.5 at$\%$Cu  identified as an orange point 1 in the phase diagram in Fig.~9. In this alloy 
crystals of primary $\alpha$-phase will begin to precipitate at 750$^\circ$C as the temperature is lowered. If cooling continues, the composition of the 
liquid will change towards the boundary curve $p_5$.
When the composition reaches the boundary curve at 605$^\circ$C (the point 2 in phase diagram), crystals of $\beta$-phase will precipitate along with crystals of $\alpha$-phase.
With further cooling, the liquid will change its composition along the boundary curve $e_3$ towards the point $U_5$ while the liquid  produce crystals of the $\gamma$-phase. 

The main interest is the four-phase reaction in the point $U_5$ (604$^\circ$C)  where crystals of $\gamma$ will begin to precipitate along with the $\beta$ 
phase while the liquid reacts with some of the crystals of the $\alpha$-phase 
\begin{equation}
\text{L} + \text{Ni}\text{Al}_3(\alpha)\rightarrow \text{Ni}_2\text{Al}_3(\beta) + \text{(Al)}(\gamma).\label{Reaction}
\end{equation}
Four various phases coexist in this point. This type of reaction (which in many ways is equivalent to the peritectic point on binary diagrams) 
is known as the tributary reaction point (because it looks like a point where two tributaries of a river meet). The product phase $\beta$ can precipitate 
along with another product phase $\gamma$ on the primary phase $\alpha$, or $\beta$ may be a potent nucleant for $\gamma$, too. The microstructure formation 
of such a ternary alloy during the directional solidification is of great interest. 

The material and model parameters considered in the simulations are listed in
Table~\ref{Tabel1}. 

\begin{table}[htbp]
\caption{Material parameters and phase-field model parameters used in the simulation.}
\label{Tabel1}
\begin{center}
\begin{tabular}{ l c}
\\ \hline  \hline
Parameter  & value used\\  \hline
$\tau$ (system time scale) & $1 \times 10^{-6}$ s\\
$l_0$ (system length scale) & 1.3$\times 10^{-8}$ m\\
$\Delta x/l_0$ (grid discretization size)	& $1$\\
$\Delta t/\tau$ (time step) & $0.025$\\
$W/l_0$ (interface width) & $1.1 $\\
$D_L^{Ni}$ (diffusion in liquid phase) & 1.2  $\times10^{-9}$ m$^2$/s\\
$D_L^{Cu}$ (diffusion in liquid phase) & 0.8  $\times10^{-9}$ m$^2$/s\\
$D_S$ (diffusion in solid phase) & $0.01 D_L$\\
$T_{U_5}$(4-phase reaction temperature) &  604 $^\circ$C\\ 
$\sigma$ (surface energy) & 0.30 J/m$^2$\\
\hline \hline
\end{tabular}
\end{center}
\end{table}

From the Gibbs free energy functions of phases we estimated the values of the equilibrium concentrations $A^{A}_{i}$, the equilibrium energies $B_{i}$
and the thermodynamic factors $X_{i}^{AB}$ at the temperature 600$^\circ$C  which is below the reaction point $U_5$. The thermodynamic parameters of the alloy system are presented in Table~2. 

\begin{table*}
 \begin{center}
\caption{Thermodynamic parameters at 600$^\circ$C used in the simulations. }
  \begin{tabular}{cccccccc}
\hline
 Phase & $A^{Ni}_{i}$,  &  $A^{Cu}_{i}$, & $B_{i}$,  & $X_{i}^{Ni}$,  & $X_{i}^{NiCu}$, & $X_{i}^{Cu}$,  & $X_{i}^{CuNi}$,\\
 & at$\%$ &   at$\%$&  J/mol-at&  J/mol-at &  J/mol-at &  J/mol-at&  J/mol-at\\ \hline
$L$ & 5.0898 & 6.5227 & -46045 &$ 4.7\cdot10^5$&$ 1.87\cdot10^5$&$ 5.0\cdot10^5$&$ 1.87\cdot10^5$\\
$\alpha $  & 25.0  &0.0& -72800& $4.0\cdot10^7$& -& $4.0 \cdot10^7$&-\\ 
 $\beta $  &  20.332 & 18.768 &-76910 & $7.1 \cdot10^5$& $6.0 \cdot10^5$&$6.3 \cdot10^5$& $6.8 \cdot10^5$\\
$\gamma $ & 0.2& 0.2& -74775  &  $7.2\cdot10 ^5$ &-& $7.2 \cdot10^5$&-\\\hline
  \end{tabular}
 \end{center}
\label{Table2}
\end{table*}

\subsection{Simulation results}

The four phase reaction were simulated at the constant temperature 600$^\circ$. Equations (\ref{PhasefieldEq0}) and (\ref{B5}) were solved numerically using the Euler method in the cubic 2D simulation box of size $200\,\Delta x$. The derivatives of the model functions $g_i$ were calculated according to Eqs.~(\ref{Deriv4}) and (\ref{Deriv5}) with $N=4$.
The simulations were started with an initial crystal of the $\alpha$-phase of radius $12\, \Delta x$. After 120 steps a nuclei of the $\beta$-phase was inserted at a random site on the solid-liquid boundary of the parent $\alpha$-phase and after 150 steps a nuclei of the $\gamma$-phase was inserted in the triple point of the $\alpha$-, $\beta$- and liquid phases.

Results of the evolution of the microstructure at various time steps are shown in Fig.~10 (a-d). Three solid phases grow from an initial multi-phase nuclei forming the two-phase boundaries. The growth velocity of the $\alpha$-phase is slower than the growth velocity of the $\beta$- and $\gamma$-phases due to the higher Gibbs free energy, so that with increasing time the product phases overgrow the crystal of the parent $\alpha$-phase. The chosen model functions serve the stability of the solution and the absence of a third phase on individual interfaces. The nucleation of the $\gamma$-phase occurs in the triple point of phases $\alpha$, $\beta$ and liquid.  No nucleation of the third phase on individual interfaces can be observed even for the larger undercooling. The lamellar-like structure forms by the overgrowing of one phase over another one.

\begin{figure*}
\label{Fig9}
\begin{center}
\includegraphics[scale=0.2]{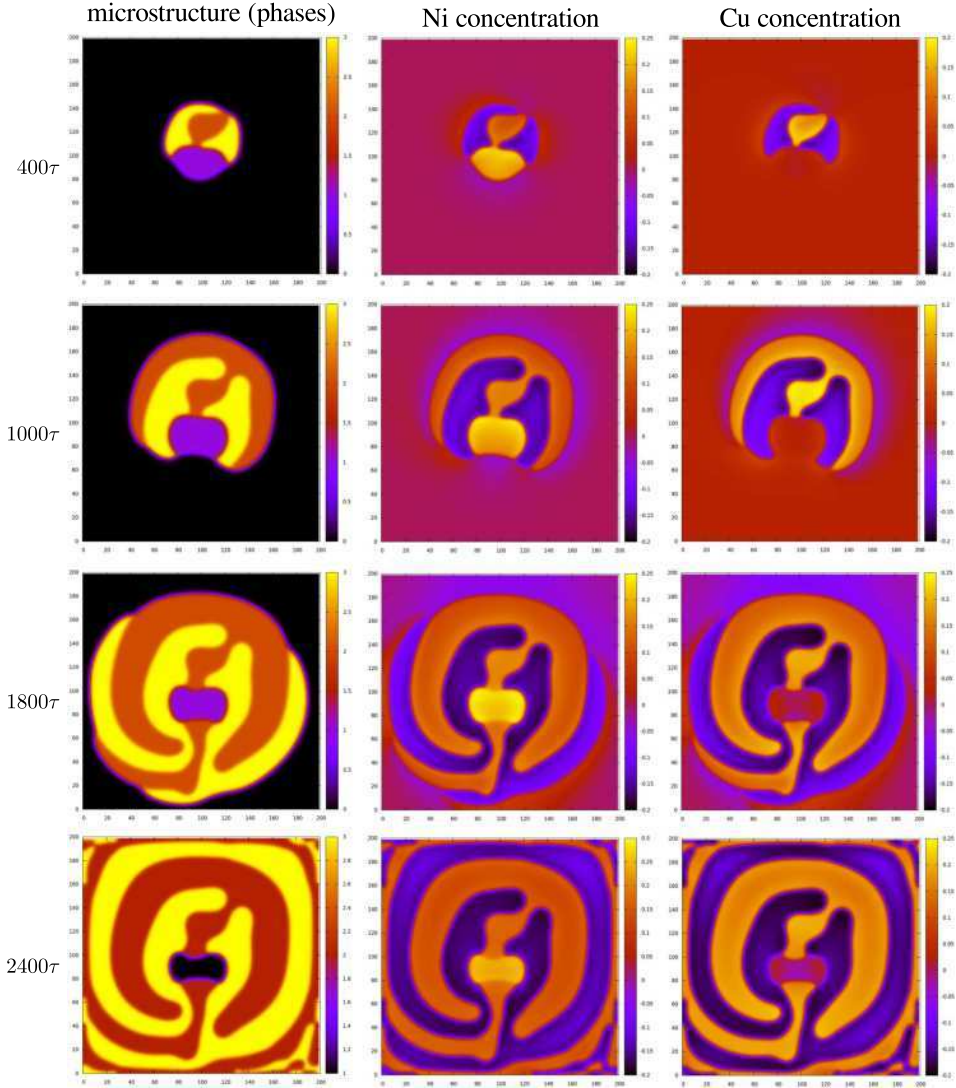}
\caption{Simulated microstructure and concentration fields evolution during isothermal four phase reaction in the 2D box without thermal noise. First column represents the microstructure at 400 (a), 1000(b) 1800 (c) and  2400 $\tau$ (d), blue area represents $\alpha$ phase, red and yellow areas represent $\beta$ and $\gamma$ phases. The corresponding concentration fields of Ni and Cu are shown in the second and third columns.}
\end{center}
\end{figure*}

 The time evolution of the Ni and Cu concentration is shown in Fig.~10 (e-l). The composition is initially homogenious until the evolution of the phase field causes the redistribution of the alloy components between the phases. It can be seen that the concentration of Ni and Cu in $\alpha$-phase is larger near the $\alpha$/$\gamma$, $\alpha$/$\beta$ boundaries and smaller at the $\alpha$-liquid boundary. This phenomenon can be clear explained by the growth of the $\alpha$-phase in the direction of the liquid-phase. The same inhomogeneities in the composition can be observed in the $\beta$-phase by the comparison of the $\beta$-liquid boundary and $\beta$-$\gamma$ boundaries.

To proof the ability of the $N$-phase model to produce the nucleation of the third phase we carried out the simulation with  an additional thermal noise \cite{Karma99} in the kinetic equation:
\begin{equation}
\xi_i=r\,\frac{RT}{H}\frac{15}{(N-1)}\sum_{k,j\neq k}^n\,p_k^2(1-p_k) p_j.
\end{equation}
where $r\in [-0.5;0.5]$ is a random value, $RT$ is the amlitude of thermal fluctuation, and $H$ is responsible for the surface energy. Due to this term a phase $i$ can nucleate heterogeneously on the interface between  phases $j$ and $k$ and grow if the energetic conditions i.e. the concnetration distribution and the surface energy are favorable.
The physical meaning is that the nucleation barrier can be overcome if the driving force is large enough.
The results of the simulation are shown in Fig.~11.
It can be observed that new thin $\beta$- and $\gamma$-phases form on the $\beta$-liquid and $\gamma$-liquid boundaries, respectively. After increasing time, if the $\alpha$ phase is closed and only $\beta$ and $\gamma$-phases grow, the resulting microstructure is similar to the eutectic lamellae structure. The same microstructure evolution were produced by the simulation of the eutectic reaction by means of the three-phase model of Folch and Plapp. The corresponding examples can be found in the works \cite{Kundin12,Ebrahimi12}.

\begin{figure*}
\label{Fig10}
\begin{center}
\includegraphics[scale=0.2]{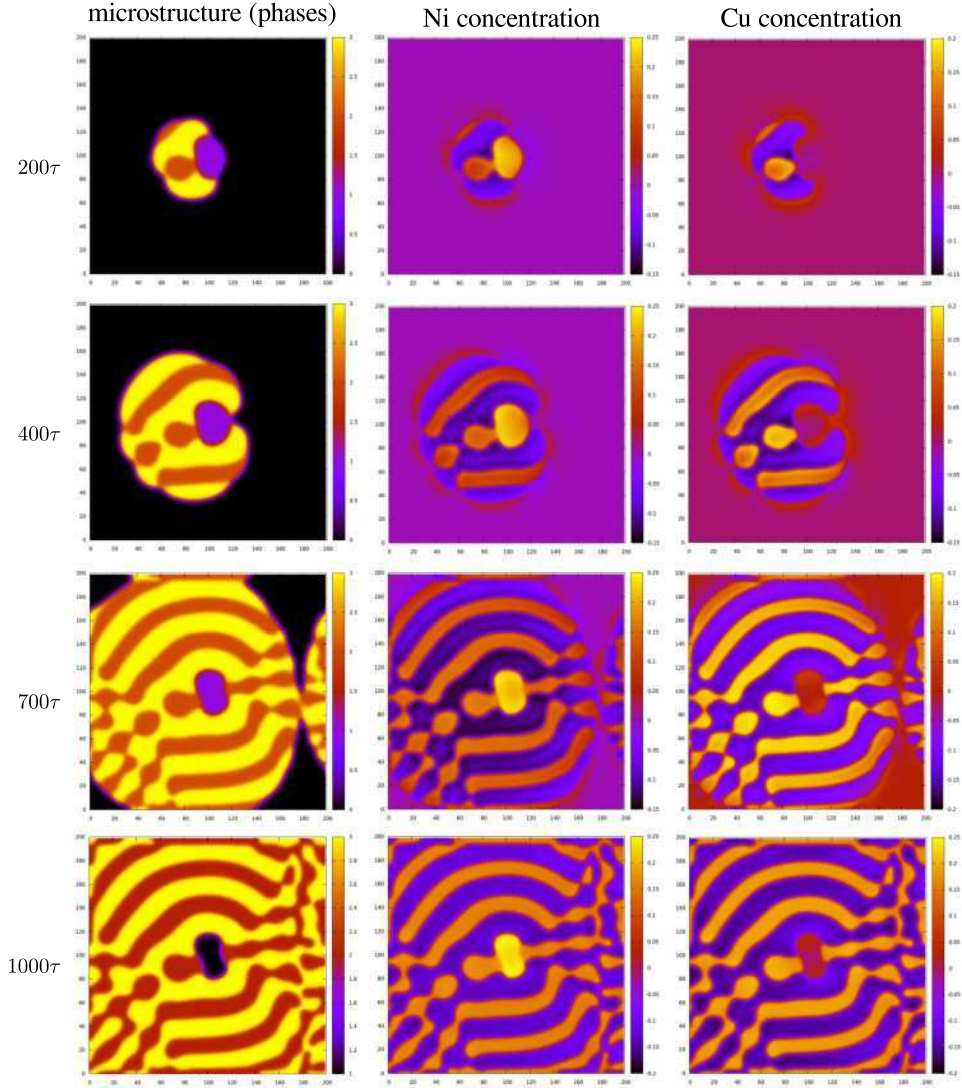}
\caption{Simulated microstructure and concentration fields evolution during isothermal four phase reaction in the 2D box with thermal noise $\xi$. First column represents the microstructure at 200 (a), 400(b) 700 (c) and  1000 $\tau$ (d), blue area represents $\alpha$ phase, red and yellow areas represent $\beta$ and $\gamma$ phases. The corresponding concentration fields of Ni and Cu are shown in the second and third columns.}
\end{center}
\end{figure*}

The comparison of the time evolution of the phase fractions for two simulated case is presented in Fig.~12. The thermal noise triggers the nucleation and produces the stable and uniform growth of the $\beta$- and $\gamma$-phases with increasing growth velocity. It can be shown in the figure that without the nucleation the lamellae of one phase overgrow the lamellae of the other phase and proceed to grow along the solid/liquid boundary where the concentration values are favorable. So far the system should wait for the moment when one phase grow enough to go around the partner phase. In the case of the additional thermal noise new thin lamellae nucleate at the the solid/liquid boundaries of the partner phase immediately after reaching the favorable concentration conditions. The amplitude of the nucleation can be adjusted in accordance to the experimental microstructure.

\begin{figure*}
\label{Fig11}
\begin{center}
\begin{tabular}{c}
\includegraphics[scale=0.45]{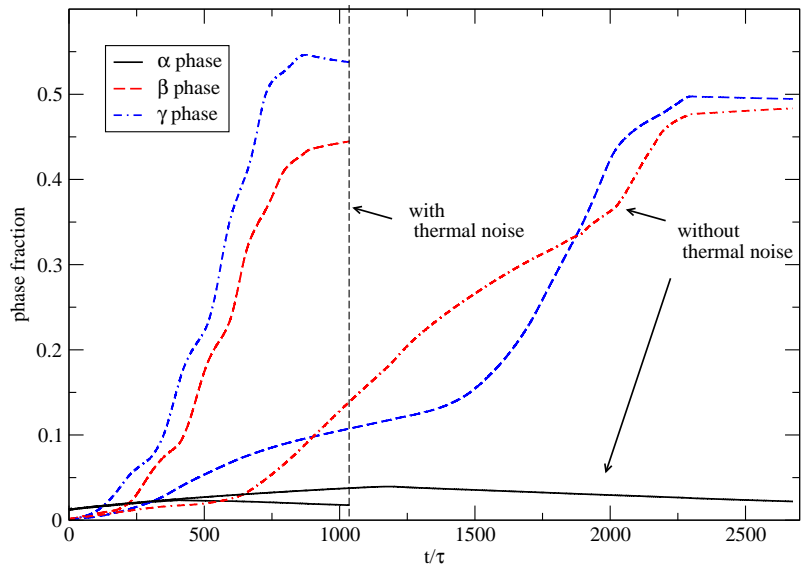}
\end{tabular}
\caption{Time evolution of the phase fractions without thermal noise and with thermal noise.}
\end{center}
\end{figure*}

Notice that we can insert a nuclei arbitrary on a solid/liquid interface and the nucleation will occur if the driving force for the nucleation is large enough. But in the present model the nucleation occurs at the right place and at the favorable energetic conditions. Moreover, the additional terms in the model allow to reduce or increase the nucleation barrier, in particular the barrier can be increased in the triple point and prevent the nucleation of the fourth phase in this points as it was shown in numerical tests above.

\section{Conclusion}

In this work we have formulated a general phase-field model in $N$-dimensional phase-field space. The main point of the model is a spatial constructed smooth total free energy functional for an arbitrary number of phases which satisfies the requirements of the stability and flatness on all individual interfaces and vertexes.  For this aim the special model functions being responsible for the interface energy barrier and for the chemical driving force of the transformation are proposed which satisfy these requirements and allow to take into account the anisotropy of the interface energies and mobility parameters.

The ability of the model to follow the Young's law and the dynamics of the system evolution is tested by the investigation of the phase-field interactions at the interfaces and the multiple junctions. It was found that the nucleation of new phases can be controlled by the additional terms in the phase-field evolution equations.
The applicability of the model to multicomponent and multiphase systems was verified by the quantitative simulation of the microstructure evolution in a ternary Al-Cu-Ni alloy in the presence of the four-phase peritectic-like reaction. It was shown that after the four-phase reaction  the three-phase reaction occures and a morphology similar to the lammelar structure is developed. Furthermore, the type of morphology and growth rate of the crystals can be controlled by the thermal noise term added to the phase-field evolution equation. 

In our future work the presented model will be used for the investugation of the microstructure evolution for various alloy compositions during cooling with various cooling rates and temperature gradients.

\section*{Acknowledgements}

The authors thanks M. Flack and D. Pilipenko from MPS of University Bayreuth for valuable discussions. The research was supported by German Research Foundation in the scope of SPP 1296.

\end{document}